\documentclass[twocolumn,showpacs,preprintnumbers,amsmath,amssymb]{revtex4}

\usepackage{graphicx}
\usepackage{dcolumn}
\usepackage{bm}

\begin{document}


\title{Correlated diffusion of colloidal particles near a liquid-liquid interface}

\author{Wei Zhang$^{1,2,\dag}$}

 \author{Song Chen$^{1}$}

 \author{Na Li$^{1}$}

  \author{Jiazheng Zhang$^{1}$}

\author{Wei Chen$^{1}$}%
 \thanks{Corresponding author.  E-mail: phchenwei@fudan.edu.cn}
 \thanks{\dag E-mail: wzhangph@gmail.com}

\affiliation{%
$^1$State Key Laboratory of Surface Physics and Department of Physics, Fudan University, Shanghai 200433, China\\
$^2$Department of Physics, Jinan University, Guangzhou 510632, China\\}

\date{April 12, 2013}

\begin{abstract}
Abstract - Optical microscopy and multi-particle tracking are used to investigate the cross-correlated diffusion of quasi two-dimensional (2D) colloidal particles near an oil-water interface. It is shown that the effect of the interface on correlated diffusion is asymmetric. Along the line joining the centers of particles, the amplitude of correlated diffusion coefficient ${D}_{\|}(r)$ is enhanced by the interface, while the decay rate of ${D}_{\|}(r)$ is hardly affected. At the direction perpendicular to the line, the decay rate of ${D}_{\bot}(r)$ is enhanced at short inter-particle separation $r$. This enhancing effect fades at the long $r$. In addition, both $D_{\|}(r)$ and $D_{\bot}(r)$ are independent of the colloidal area fraction $n$ at long $r$, which indicates that the hydrodynamic interactions (HIs) among the particles are dominated by that through the surrounding fluid at this region. However, at short $r$, $D_{\bot}(r)$ is dependent on $n$, which suggests the HIs are more contributed from the 2D particle monolayer self.

\end{abstract}

\pacs{82.70.Dd, 68.05.Gh, 05.40.-a, 83.85.Jn}
\maketitle


\section{introduction}

Much attention has been attracted to the dynamic behavior of confined colloidal suspensions recently \cite{1,2,3,4,Levine02,5}. In real circumstances, particles are usually spatially confined in special environments, such as microfluidic devices, porous media, fluid interface and cell membrane \cite{6,Sickert03,Chen08,Ortega10,Pralle00}. The dynamical behaviors of colloids in spatially confined environments are more complicated than that in unbounded three-dimensional (3D) fluid bulk. In the unbounded 3D bulk suspensions, the longitudinal and transverse correlated diffusion coefficient $D_{\|}$ and $D_{\bot}$ are well known as $D_{\|}\varpropto 1/r$, $D_{\bot}\varpropto 1/r$ and $D_{\|}=2D_{\bot}$ \cite{Crocker00}. For the particles confined near the solid wall, the strength of hydrodynamic interactions (HIs) decay with inter-particle separation $r$ as $1/r^2$ \cite{4,Cui04,Diamant05}. In addition, spatial symmetry break due to the boundary conditions brings asymmetry feature to cross correlated diffusion of particles \cite{2,Oppenheimer09,Cui04,Diamant05,Crocker00}. For the particles on the air-water interface or a viscous membrane, both experimental \cite{Prasad06} and theoretical \cite{Oppenheimer09} studies show $D_{\|}\varpropto 1/r$ and $D_{\bot}\varpropto 1/r^2$. Oppenheimer's $et$ $al.$ calculation \cite{Oppenheimer10} showed that at the membrane adjacent to a solid wall, the correlated diffusion of particles is a function of the distance between the membrane and solid wall.

The influence of different boundary conditions could be directly reflected on diffusion behavior of particles \cite{Perkins91,Bickel07,Cichocki07,Blawzdziewicz10,Lee79}. Many studies base on solid wall condition, a non-slip boundary, which could cut off the fluid field in its vicinity. For a fluid-fluid interface (say a soft wall), there is a slip boundary which partially transforms the flow field around. Compared with extensive investigations on the solid wall's effects, there are few experimental studies about the influences of the fluid-fluid interface on particles' correlated diffusion.

In this paper, we report the experimental measurement of the cross-correlated diffusion of colloidal particles near the water-decahydronaphthalene (decalin) interface. The particle monolayer is similar to a viscous membrane. The correlated diffusion of particle pair shows asymmetric behavior in two normal directions. Along the line joining the centers of particles, the decay rate of correlated diffusion is constant. At the perpendicular direction, the decay rate increases with the particle separation. In both directions, the correlated diffusion is independent of the area fraction of particles for long particles separation. The influence of fluid-fluid interface on the dynamic behavior of the colloidal monolayer tends to be saturated, when the particle separation is much larger than the distance between the interface and monolayer. These phenomena reveal weights of HIs through different paths.

The remainder of the paper is organized as follows. Section II is devoted to the description of the experimental setup. Experimental results and discussions are presented in Sec. III . Finally, the work is summarized in Sec. IV.

\section{Experimental setup}
Two kinds of colloidal particle with the same diameter d =2.0v$\mu m$ are used in the experiment. The first one is polystyrene (PS) latex spheres purchased from Invitrogen with sulfate and high density of carboxyl functional groups on the surface. The other one is silica spheres purchased from Bangs, which have anionic $SiO^-$ groups on the surface when dispersed in water. The density of the silica spheres are about 1.9 times larger than that of PS. Both the polystyrene and silica spheres are the representative of charged particles commonly used in colloidal science.

The experimental setup is similar to that described in Ref. \cite{Peng09}. The sample cell is made from a stainless steel disk with an inner container in which there is a hole with the diameter 8.3 $mm$. The bottom of the hole is sealed with a 0.1 $mm$ thick glass cover slip, which also serves as an optical window. We first fill the hole to the top edge with deionized water of 18.2 $M\Omega\cdot cm$ in which the particles cleaned 7 times by centrifugation were suspended. And then we add decalin (a mixture of cis and trans with density 0.896 $g/cm^3$) to the top of the water, filling the entire height of the inner container. The solvent decalin was purchased from Sigma-Aldrich. Another cover slip is used to cover the top of the inner container with the water-decalin interface sandwiched between the top and bottom cover slips. Then we overturn the cell. Particles fall onto the water-decalin interface caused by the gravity for one or two hours and form a monolayer, as shown in Fig. 1. The colloidal particles are hydrophilic and will prefer to stay in water rather than in decalin. In addition, particles keep a certain distance away from the interface because of image charge in decalin. The image charge force is repulsive for particles approach the decalin-water interface from the medium (water) with the higher dielectric constant \cite{Mbamala02}. The water and decalin layers are both 0.8 $mm$ high. The distance between the cover slip and colloidal monolayer is long enough to eliminate the influence of cover slips.

 With an inverted microscope Olympus CoolSNAPIX71, the motion of colloidal particles was recorded by a digital camera (Prosilica GE1050) of 14 frames per second. Each image sequence includes 500 consecutive frames ($\simeq 35.7$ sec). Each image is of size $176*176 \mu m^2$. From these image sequences, using homemade software, we obtain the particle positions and construct trajectories with a spatial resolution of 60-100nm.

\section{Experimental results and discussions}
Interaction potential $U(r)$ of particles was estimated from pair correlated function $g(r)$ of particles. $r$ is the particle separation. At dilute limit, $g(r)$ is related to $U(r)$ through the function of $U(r)/k_B T=-ln(g(r))$. In our experiment, $g(r)$ was calculated by averaging $10^6$ particle positions under area fraction $n=0.04$ for PS and silica spheres.  For even higher area fractions $n=0.08$, measured $g(r)$ of PS and silica spheres could overlap with the formers respectively. That indicates the area fraction used could be regarded as the dilute limit. Curves of $U(r)$ are shown in Fig. 2. Both particles look like hard spheres while $r>2.5d$. When $r$ is shorter than $2.5d$, an attractive potential well appears. For silica particles, the depth of attractive potential well is around $0.3 k_B T$ at $r=1.3d$. For PS particles, the depth is around $0.4 k_B T$ at $r=1.6d$. What we found accords with the results in Ref. \cite{Grier03,Peng05}: like charge particles could attract each other near an interface. PS particle shows a wider and deeper attractive potential than silica, because it has more surface charge than silica \cite{Peng05}.

PS monolayer is located at higher position above the oil-water interface than silica because PS sphere has less mass density and more surface charge. The average distance $z$ from particle monolayer to the interface could be estimated by measuring the single diffusion coefficient of particles.  From the particle trajectories ${\bf s}(t)$ , we calculated the single-particle mean square displacement (MSD) $\langle \Delta {\bf s}^2(\tau) \rangle = \langle |{\bf s}(t+\tau) - {\bf s}(t)|^2 \rangle $. The self-diffusion coefficient $D_S$ is obtained according to the equation $\langle \Delta {\bf s}^2 (\tau) \rangle =4D_S* \tau$ for different particle area fraction $n$. Figure 3 shows the dependence of the diffusion coefficient $D_S/D_0$ on $n$ for PS and silica spheres. $D_0=k_BT/3\pi \eta_w d$ is the diffusion coefficient of a single particle in water.  The solid lines in Fig. 3 are the second-order polynomial fitted results,
 \[
 D_S/D_0=\alpha(1-\beta n-\gamma n^2).\qquad\qquad  (1)
\]
The parameter $\alpha$ stands for the local viscosity felt by a single sphere at dilute limit: the larger $\alpha$ is , the smaller the viscosity is. The parameter $\beta$ reflects the strength of two body HIs between the spheres: the smaller $\beta$, the smaller two-body HIs. The fitted values of $\alpha$, $\beta$ and $\gamma$ are given in Table I.

The value of $\alpha$ can be used to estimate the distance between the interface and the sphere monolayer $z$. According to the classical prediction given by Lee $et$ $al.$ \cite{Lee79,Wang09}, the distance from particle's center to the interface $z$ can be written as:
 \[
 \frac {z} {a}=\frac {3(2\eta_w-3\eta_o)} {16(\eta_w+\eta_o)(\alpha-1)},\qquad\qquad  (2)
\]
where $a$ is particle radius, $\eta_w$ is the viscosity of water and $\eta_o=2.5$ $cP$ is the viscosity of decalin at 22.5 $^0C$. Substituting the fitted values of $\alpha$ into Eq. 2, we obtain $z=2.3\mu m\pm0.2\mu m$ ($1.4\mu m\pm0.1\mu m$) for PS (silica) sphere. The calculation is consistent well with the experimental observation under the microscope: particles keeps $1-2 \mu m$ distance away from the interface.

The particles slightly fluctuate in vertical direction under $k_B T$. The fluctuation amplitude of silica particles is smaller than PS because the former has heavier mass density. Within the area fraction region we measured, particles near the interface could be viewed as a monolayer. We focus our investigation on the lateral motion of particles, parallel to the oil-water interface, in following part of the paper.

Following the trajectory ${\bf s}^i(t)$ of individual particles $i$, particles' cross-correlated motion is obtained via the ensemble averaged tensor product of the particle displacements \cite{Crocker00}:
 \[
M_{x y}(r,\tau)=<\Delta s^i_x(t,\tau)\Delta s^j_y(t,\tau)>_{i \neq j,t}. \qquad  (3)
\]
Here $ \Delta s^i_x(t,\tau) = s^i_x(t+\tau) - s^i_x(t) $. $i,j$ are particle indicates, $x$ and $y$ represent different coordinates, and $r$ is the separation between particle $i$ and $j$ (shown as the inset of Fig. 4). The off-diagonal elements are uncorrelated. We focus on the diagonal elements of this tensor product: $M_{rr}$ which indicates the correlated motion along the line joining the centers of particles (called parallel direction), and $M_{\theta\theta}$ which represents the correlated motion perpendicular to this line (called perpendicular direction). We found that the measured correlated motion $M_{rr}$ and $M_{\theta\theta}$ are linear functions of $\tau$ for small lag time $\tau$. Thus, the cross-correlated diffusion coefficients are defined as $D_{\|}=M_{rr}/2\tau$ and $D_{\bot}=M_{\theta\theta}/2\tau$.

Figure 4 exhibits $D_{\|}$ and $D_{\bot}$ of PS and silica spheres as a function of $r$ for different area fraction $n$. Each curve of an arbitrary $n$, whose deviation was given in the legend of Fig.4, was obtained by averaging at least $10^6$ particle positions. The behaviors of $D_{\|}$ and $D_{\bot}$ of PS and silica monolayer are qualitatively similar. With the increase of $r$, $D_{\|}$ and $D_{\bot}$ of PS and silica sphere decrease following the power law. The specific form of the power law will be discussed in the following section. When $r>2.5d$, the curves of $D_{\|}(r)$ and $D_{\bot}(r)$ for different $n$ almost collapse onto a single curve respectively, i.e., particle area fraction hardly affects the particles' cross-correlated motion. When $r<2.5d$, however, $D_{\bot}(r)$ is dependent on the area fraction $n$. With the increase of $n$, the decay rate of $D_{\bot}(r)$ decreases, i.e., the curves become more flat.

The mechanism of $n$ independence effect mentioned above is in line with Oppenheimer $et$ $al.$'s theoretical description in Ref. \cite{Oppenheimer10}. In the particle monolayer, the far-field response of the correlated particle motion mainly arises from the momentum diffusion through the 3D surrounding fluid. Usually the viscosity of 3D fluid suspension is the function of volume fraction $\phi$ of particles. In our system $\phi$ could be regarded as zero almost, no matter how the area fraction $n$ of monolayer changes.  The viscosity of 3D fluid almost keeps as a constant. Thus the far-field 3D HIs between the particles are independent of the area fraction $n$, so does the cross-correlated diffusion $D_{\|}(r)$ and $D_{\bot}(r)$. Cui $et$ $al.$ \cite{Cui04} also observed the similar concentration independent effects for particles confined between two plates, but the reasons to their phenomena are different. In the confined particle monolayer, the fluid momentum is absorbed by the solid boundaries, and the far-field fluid response arises solely from mass-dipole perturbation, which is not influenced by the presence of neighbor particles.

The situation will be more complicated for short $r$ ($<2.5d$), because the particles reveal an attractive potential in the range (Fig. 2). Both HIs and thermodynamic interaction involve in particle's correlated motion now. For the potential is attractive, the correlation of particle motion will become stronger in the range. As a result, the curves become more flat. For HIs, the near-field response of the correlated motion mainly arises from the momentum diffusion through the two-dimensional (2D) colloidal layer itself. The 2D monolayer viscosity $\eta_m$ increases with the area fraction $n$ of the particles. As a result, the correlated motion is a function of $n$.

To focus on the influences of the oil-water interface on the correlated diffusion and eliminate the $n$ effects, we average the data of different $n$ in Fig. 4. Figure 5 shows the averaged correlated diffusion coefficient $D_{\|}(r)$ and $D_{\bot}(r)$ normalized by $\alpha D_0$, which is the single particle diffusion coefficient obtained by Eq. 1, as a function of particle separation $r$. At the parallel direction, there is $D_{\|}(r)/ \alpha D_0=$ $A_{\|}/r^{\lambda_{\|}}$. Here, $A_{\|}$ is the amplitude coefficient and $\lambda_{\|}=1$ is the decay rate of cross-correlated motion. The interface affects only the amplitude $A_{\|}$ but not the decay rate $\lambda_{\|}$. The amplitude $A_{\|}$ of the silica spheres is larger than that of PS, which is in accordance with the results from single diffusion measurement: the fitted value of $\beta$ of silica is greater than that of PS in Table I. The phenomenon indicates that the strength of HIs between the particle pair is more strong in the silica monolayer than in PS monolayer. The amplitude of $A_{\|}$ of the silica spheres is found almost 1.2 times larger than that of PS, which number is just equal to the ratio of the $\beta$ ($0.78/0.62=1.2$). The same decay rate $\lambda_{\|}=1$ for both monolayers indicates that the form of HIs is hardly affected by the interface in the parallel direction.

At the perpendicular direction, $D_{\bot}(r)/ \alpha D_0=$ $A_{\bot}/r^{\lambda_{\bot}}$. Both the amplitude $A_{\bot}$ and decay rate $\lambda_{\bot}$ are functions of the distance $z$. The decay rates $\lambda_{\bot}$ of the silica spheres is larger than that of PS at short $r$ (1.8 vs. 1.5, as labeled in Fig. 4). Different from the constant ${\lambda_{\|}}$, the decay rate $\lambda_{\bot}$ increases with $r$. At the long $r$ limit, both $\lambda_{\bot}$ in PS and silica monolayers merge into a value of around 2, as shown in Fig. 5. The inset of Fig. 5 shows that the scaled diffusion coefficient $\widetilde{D}_{\bot}(\widetilde{r})=D_{\bot}(\widetilde{r})/ (\alpha \beta D_0 $) for both PS and silica particles, where $\widetilde{r}=r/z$. Both curves are overlapped at long region of $\widetilde{r}>15$, which indicates the effect of the liquid-liquid interface tends to be saturated.

We compare our results with the studies before for different boundary conditions: the colloidal spheres dispersed in unbounded 3D bulk, confined by the solid wall or at the air-water interface \cite{Cui04,Diamant05,Prasad06,Oppenheimer09}. We find the cross correlated motion of colloidal spheres near oil-water interface is similar to that of a viscous membrane \cite{Prasad06,Oppenheimer09}: $\lambda_{\|}=1$ and $\lambda_{\bot}=2$ at long $r$ limit. For a viscous membrane, long $r$ limit means that $r \gg L_s$, where $L_s \equiv \eta_m /2 \eta_b$ is Saffman length \cite{Saffman76}. $\eta_m$ is the 2D membrane viscosity and $\eta_b$ is the 3D bulk viscosity of surrounding fluid. For our system, limit of long $r$ means that $r\gg Max \{z, L_s\}$. $Max \{z, L_s\}$ is the larger one between $z$ and $L_s$.

Theory in \cite{Oppenheimer09} shows that the $D_{\|}(r)/ \alpha D_0$ does not depend on membrane viscosity $\eta_m$ for large $r$, but $D_{\bot}(r)/ \alpha D_0$ is dependent on $\eta_m$. This prediction accords with our results of $\lambda_{\|}$ and $\lambda_{\bot}$. Our results suggest that the colloidal monolayer near the the interface could be regarded as a 'membrane'. Even though the precise value of 2D viscosity $\eta_m$ of colloidal monolayer is hard to be obtained from our present data, the value of $\lambda_{\bot}$ (1.8 or 1.5) indicates the $\eta_m$ usually is very small. For a 'membrane' with viscosity $\eta_m$, the value of $\lambda_{\bot}$ will approach to 2 at low $\eta_m$ limit and approach to 0 for high $\eta_m$ limit \cite{Prasad06}. Precisely, $D_{\bot}(r)$ will tend to a logarithmic form at high $\eta_m$ limit. Hence $L_s$ should be very small also, and the limit of long $r$ usually means $r\gg z$ in our system.

At first glance, the result that $\lambda_{\bot}=1.8$ for the silica spheres is larger than $\lambda_{\bot}=1.5$ for PS seems confusing, which conflicts with the fact that a larger $\lambda$ should corresponds to a smaller monolayer viscosity. Since the silica spheres are closer to the interface, it is located in a more viscous environment comparing with PS. Thus, $\lambda_{\bot}$ of the silica spheres seems should be less than that of PS. The fact is that the viscosity $\eta(z)$ felt by the spheres is the 3D viscosity of the local surrounding fluid, not the 2D viscosity $\eta_m$ of the membrane. The influence of this surrounding viscosity $\eta(z)$ has been removed by $1/(\alpha D_0)$ scaling, where $\alpha$ stands for the effect of local viscosity $\eta(z)$. The result that $\lambda_{\bot}$ of silica is larger than $\lambda_{\bot}$ of PS stems from HIs modified by the interface. Oil is more viscous than water. The flow field induced by the particle in water is suppressed by the oil-water interface. HIs between particles decay faster with their separation $r$ when the monolayer is closer to the interface, which leads to a larger measured $\lambda_{\bot}$ for a smaller $z$.

Similar to the mechanism of a membrane near a solid wall \cite{Oppenheimer09}, HIs among the particles in our system transmit through three paths: I) The 2D flow through the monolayer. II) The flow through the fluid layer sandwiched between the monolayer and interface. III) The flow through the upper water bulk or the below oil bulk.  HIs through path I is dependent on 2D monolayer viscosity $\eta_m$ , while HIs through path II and III are dependent on 3D viscosity $\eta_o$ and $\eta_w$ of surrounding fluid.

The cross-correlated motion caused by HIs through Path I is a function of $n$, because $\eta_m$ is a function of $n$ usually. HIs through Path I only contributes to correlated motion when the separation $r$ is within the order of Saffman length $L_s =\eta_m/(\eta_o+\eta_w)$. In our system $\eta_m$ of the monolayer is very small. With $r$ increasing, the weight of HIs through path I decreases quickly. Meanwhile, HIs through Path II contributes to correlated motion all the time. So, correlated motion is a function of $z$ (see Fig. 5). At long separation limit $r\gg Max \{z, L_s\}$, the HIs through path III dominate. Since $\eta_o$ and $\eta_w$ are hardly changed with area fraction $n$, the correlated motion should be independent of $n$ (see Fig. 4). The correlated motion is also independent of $z$ for $r\gg z$. At middle range of $r$, HIs through all three paths contribute, their relative weights change gradually with $r$. The decay rate of the cross-correlated motion shows a crossover tendency.

\section{Conclusion}
We investigated experimentally the cross-correlated diffusion of colloidal particles near the oil-water interface. Correlated diffusion coefficient $D_{\|}(r)$ and $D_{\bot}(r)$ are independent of area fraction $n$ unless at the short pair separation $r$. The interface affects correlated diffusion coefficient $D_{\|}(r)$ and $D_{\bot}(r)$ distinctly. The interface enhances the amplitude of $D_{\|}(r)$, but does not affect its decay rate. The results indicate: along the line joining the centers of particle pair, the interface modifies the amplitude of the HIs between the particle pairs, but not the form of HIs. At the direction perpendicular to the line, the influence of the interface changes with increasing $r$. The interface enhances the decay rate at short $r$, while such influence tends to be saturated at the long $r$ limit ($r\gtrsim 15z$, $z$ is the distance between the interface and the monolayer). The dependence of the decay rate on $r$ evidences that the weights of HIs through different paths change gradually.

\acknowledgements
We thank for helpful discussions with Penger Tong from HKUST. W. Chen and J. Zhang acknowledge National Science Fund for Talent Trainning in Basic Science (Grant No.J1103204).

\[
\
\]

Figure captions:

Fig. 1. (color online) (a) Schematic view of the system. (b) Optical microscope image of silica spheres ($d=2.0\mu m$) suspended near the water(up)-decalin(down) interface at area fraction $n=0.18$.

\[
\
\]

Fig. 2. (color online) Measured interaction potential $U(r)$ for PS (solid circle symbols) and silica spheres (open square symbols). Area fraction $n=0.04$ for PS and silica particles.

\[
\
\]

Fig. 3 (color online) Measured self diffusion coefficient $D_{S}$ scaled by $D_0$ as a function of particles' area fraction $n$. Different symbols represent the data for different particles. The solid lines show the second-order polynomial fitting $D_S/D_0=\alpha(1-\beta n-\gamma n^2)$.

\[
\
\]

Fig. 4 (color online) Measured correlated diffusion coefficient $D_{\|}$ (solid symbols) and $D_{\bot}$ (open symbols) as a function of inter-particle distance $r$ with various values of the area fraction $n$ for (a) PS and (b) silica spheres. The solid lines with the slope $-1$, $-1.5$ and $-1.8$ are the guides for eyes. The inset shows the geometry of measuring correlated diffusion.

\[
\
\]

Fig. 5 (color online) The mean correlated diffusion coefficient $D_{\|}/ \alpha D_0$ (solid symbols) and $D_{\bot}/ \alpha D_0$ (open symbols) obtained by averaging the data for various values of the area fraction $n$ in Fig. 4. The square and circle represent silica and PS particles, respectively. The inset shows the correlated diffusion coefficient $\widetilde{D}_{\bot}$ (open symbols) as a function of scaled inter-particle distance $\widetilde{r} (=r/z)$ for PS and silica spheres. The solid lines with the slope $-1$ and $-2$ are the guides for eyes.

\begin{table}
\caption{\label{tab:table2} The distance $z$ from particle's center to the interface and the fitted values of $\alpha$, $\beta$ and $\gamma$ in Fig. 3. Here $a=1\mu m$ is the radius of particles.}
\begin{ruledtabular}
\begin{tabular}{cccccccc}
 Sample  &  $z/a$  & $\alpha$ & $\beta$  & $\gamma$\\
\hline
PS & 2.3 $\pm$ 0.2 & 0.87 $\pm$ 0.01 & 0.62$\pm$ 0.12 & 1.54$\pm$ 0.61  \\
Silica & 1.4 $\pm$ 0.1 & 0.78 $\pm$ 0.01 & 0.78$\pm$ 0.02 & 0.76$\pm$ 0.03  \\
\end{tabular}
\end{ruledtabular}
\end{table}

\newpage

\begin{figure}
\includegraphics[height=24cm]{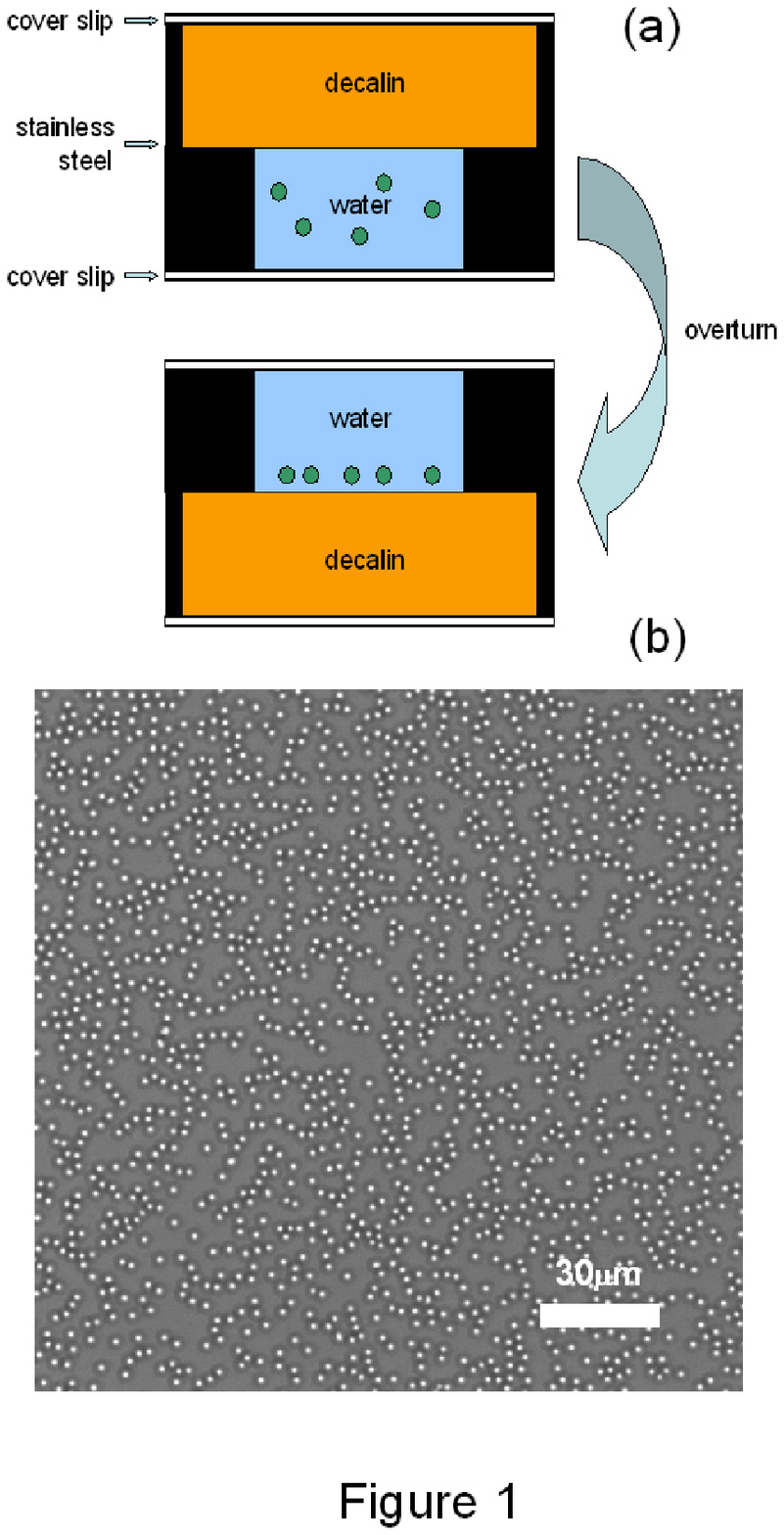}
\end{figure}

\begin{figure}
\includegraphics[height=12cm]{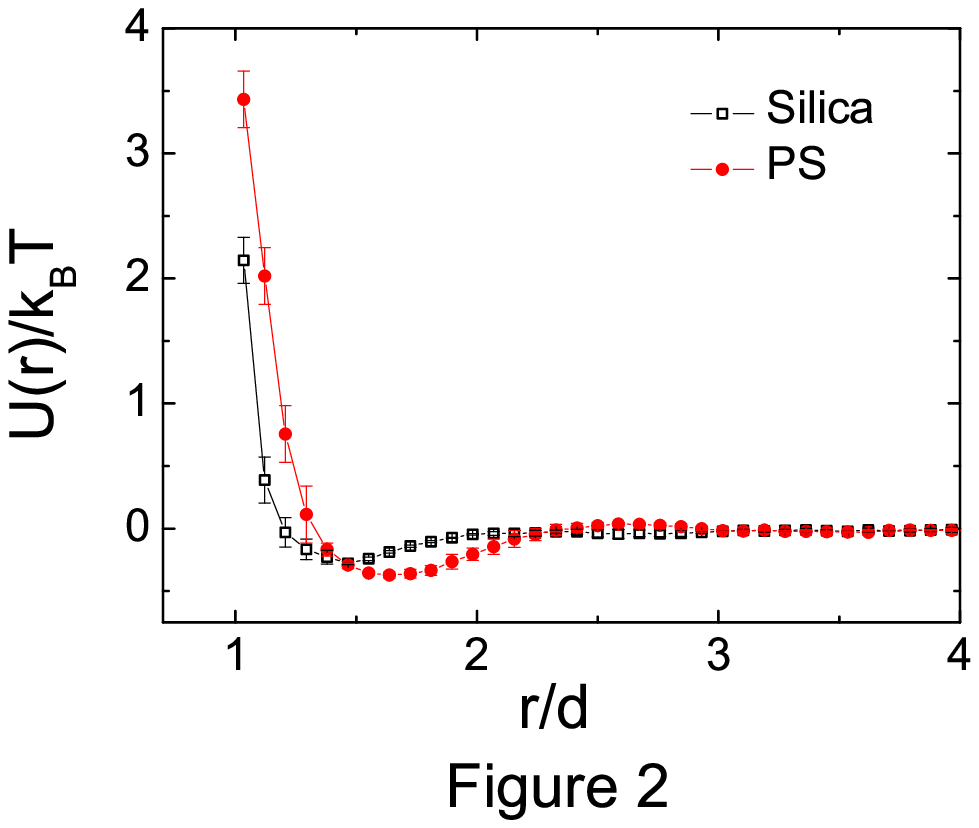}
\end{figure}

\begin{figure}
\includegraphics[height=12cm]{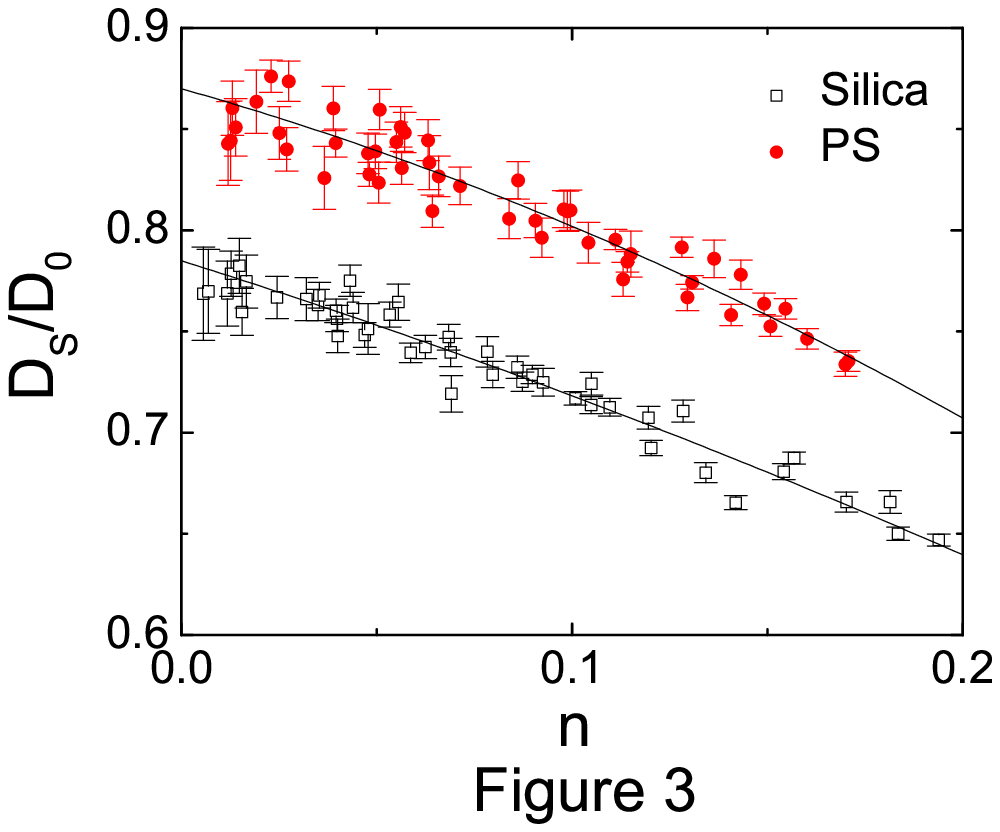}
\end{figure}

\begin{figure}
\includegraphics[height=24cm]{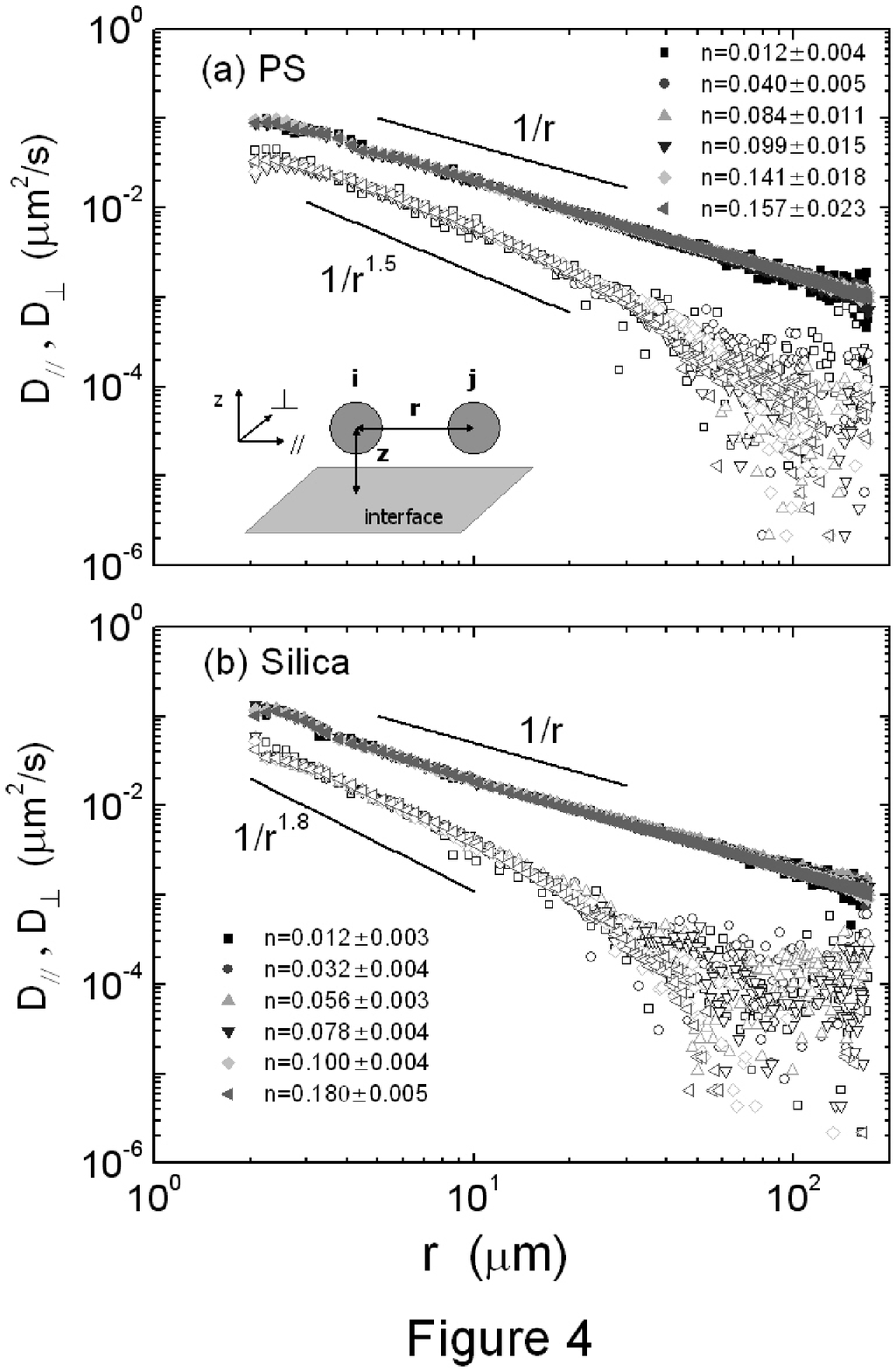}
\end{figure}

\begin{figure}
\includegraphics[height=12cm]{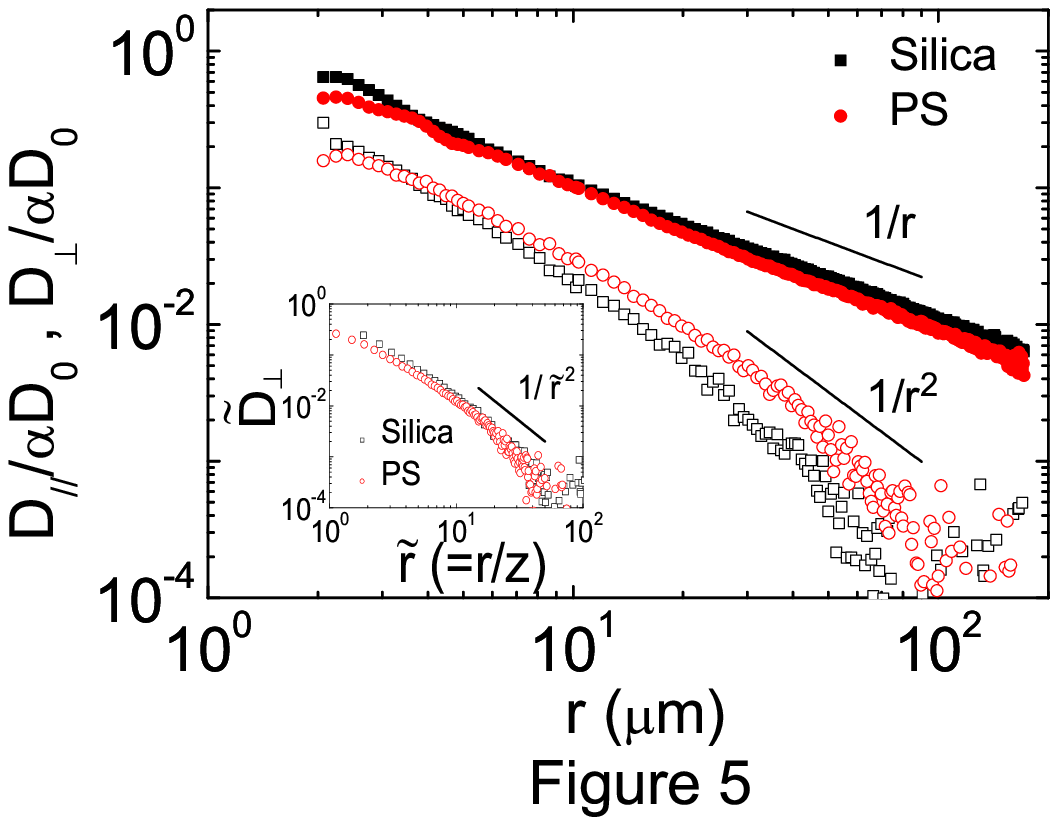}
\end{figure}

\end{document}